REVIEW ARTICLE

# Advancements in Hematology Analyzers: Next-Generation Technologies for Precision Diagnostics and Personalized Medicine

Aahsan Iqbal[a], Sohail Khalid[b] and Mujeeb ur Rehman[c]

[a]Biomedical Engineering Department, Riphah International University, Islamabad, Pakistan; [b]Electrical and Computer Engineering Department, Riphah International University, Islamabad, Pakistan; [c]School of Computer Science and Informatics, De Montfort University, Leicester, UK

ABSTRACT
Hematology analyzers are essential diagnostic and monitoring tools for detecting blood diseases. Although contemporary analyzers produce only basic insights, they are often not as detailed as required under the personalized medicine paradigm. Next-Generation Hematology Analyzers (NGHAs) are revolutionary newcomers in the field, with significant advantages over regular hematology analyzers. They provide deeper insights into cellular morphology, function, and genetic profiles. This detailed information opens up possibilities for tailor-made diagnostic and therapeutic approaches in precision medicine. This review presents some revolutionary technologies that have changed hematology analyzers and provides an overview of their limitations, basic functions, and influence on clinical practice. It focuses on the integration of state-of-the-art technologies, such as microfluidics, advanced optics, artificial intelligence, flow cytometry, and digital imaging, empowering NGHAs to improve diagnostic accuracy, rapidly detect diseases, and support flexible, targeted therapy. Hints regarding point-of-care hematology testing are also provided to discuss its implications for transforming healthcare patterns. This review highlights the data management, standardization, regulatory, and ethical challenges associated with these technologies. A review tracking the current state-of-the-art and trends for the future is provided to show how these advancements may reconfigure hematology analyzer design and act as a stepping stone for future therapeutic reforms.

## 1. Introduction

Blood analysis remains one of the most widely used techniques in clinical diagnosis, which helps to gain insights into several physiological functions that reflect possible disease states. A hematology analyzer performs automated blood cell count, size, and morphology measurements quickly and fairly efficiently; hence, it is an indispensable and critical diagnostic tool in healthcare Chhabra (2018). However, most mainstream

CONTACT A. Iqbal. Email: aahsan.iqbal@systechs.pk
CONTACT S.Khalid. Email: s.khalid@riphah.edu.pk
CONTACT M.Rehman. Email: mujeeb.rehman@dmu.ac.uk

Table 1.: General comparison between conventional hematology analyzers and NGHAs

| Conventional Hematology Analyzer | Next Generation Hematology Analyzer |
| --- | --- |
| 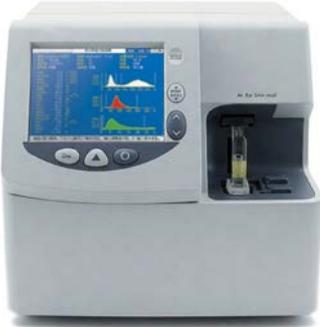 | 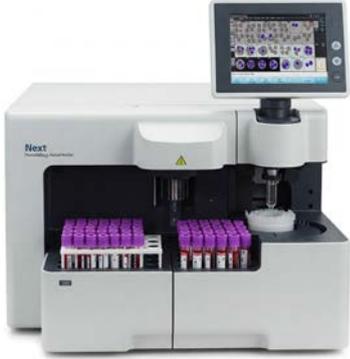 |
| <ul><li>**Measurement:** Impedance and optical light scattering-based measurement</li><li>**Throughput:** Low to moderate sample throughput</li><li>**Cell Counts:** Limited basic set of cell counts</li><li>**Loading:** Mainly manual sample loading</li><li>**Reports:** Produces basic numerical reports</li></ul> | <ul><li>**Measurement:** Advanced flow cytometry, HD digital imaging, AI-powered analysis</li><li>**Throughput:** High to very high sample throughput</li><li>**Cell Counts:** Expanded cell counts with visual cell morphological presentation</li><li>**Loading:** Full automation in sample loading and handling</li><li>**Reports:** Produces detailed graphical reports</li></ul> |

analyzers are limited in terms of the number of measured parameters, which provides a narrow understanding of blood cell health. Conventional analyzers predominantly use impedance and light-scattering techniques to identify and count blood cells Vembadi, Menachery, and Qasaimeh (2019), a method that is satisfactory for the easy enumeration of cells, but cannot yield all pertinent information inside the cell, which is required for precision approaches to medicine. There are exciting new areas in the field of precision medicine that demand tools to tailor treatment to an individual's unique biology. To achieve this, a thorough understanding of blood cells, their shape and function, and their genetic makeup is required. Next-generation hematology analyzers have come into play. These cutting-edge devices are designed to probe much deeper into the blood cells and spur all the possibilities that lie ahead in personalized medicine. Table 1 presents hypothetical depictions of conventional and NGHAs, and highlights the major differences between the two technologies. This article outlines the considerations concerning the development and operation of such instruments. These aspects are expected to influence future biomedical engineering applications.

## 2. Research Methodology

The review study protocol is described in the following five subsections.

## 2.1. Defining the Scope of the Review

This review explores recent advancements in hematology analyzers and their impact on clinical diagnostics and personalized medicine. It synthesizes current knowledge on next-generation analyzers while focusing on how these technologies are transforming the detection, diagnosis, and monitoring of hematological disorders. This review addresses the need for improved diagnostic accuracy and faster turnaround times in hematology. Traditional methods often lack sensitivity for early disease detection and tailored interventions, thereby affecting patient outcomes and healthcare efficiency. The increasing complexity of hematological disorders necessitates the development of more sophisticated diagnostic tools. This review focuses on automated hematology analyzers in clinical and research settings. The scope includes AI and machine learning for cell morphology and automated differential counting; advanced flow cytometry for immunophenotyping and rare cell detection; high-resolution digital imaging; and point-of-care testing. The context is the evolving landscape of clinical hematology in which precision diagnostics and personalized treatment drive innovation. This review considered studies published mainly within the last decade; however, older studies that are most relevant to this review are also not overlooked.

## 2.2. Search strategy

This review employed a systematic search strategy to identify relevant literature on advancements in hematology analyzers. The primary databases used in this study are PubMed, IEEE Xplore, Scopus, and Google Scholar. These databases were chosen for their comprehensive coverage of biomedical and scientific literature. The search terms included various combinations of the following keywords: *hematology analyzer*, *automated hematology*, *flow cytometry*, *digital imaging*, *artificial intelligence*, *machine learning*, *point-of-care testing*, *hematological disorders*, *precision diagnostics*, and *personalized medicine*. Boolean operators (AND, OR, and NOT) were used to refine search results. Filters were applied to limit the results to English language publications and studies published within the last decade mainly (2014-2024), ensuring that the review focused on current advancements. This timeframe was selected to capture the most recent technological developments in the field. Table 2 lists the queries used to search the relevant articles in different databases, and Figure 1 depicts their distribution. In addition to the database searches, a manual search of relevant journals and reference lists of identified articles (snowball sampling) was conducted to identify any potentially missed publications, including those published in earlier timeframes. While gray literature was not formally searched, relevant conference proceedings identified through database searches were considered.

## 2.3. Inclusion and Exclusion Criteria

Specific inclusion and exclusion criteria were established to ensure selection of high-quality studies. These criteria were applied systematically to minimize bias and ensure that the review focused on the most pertinent research.

### 2.3.1. Inclusion Criteria

The inclusion criteria for this study were as follows:

Figure 1.: Distribution of articles

Table 2.: Search queries used in various databases for the study.

| Database | Search Query |
|---|---|
| PubMed | ("hematology analyzer"[All Fields] OR "automated hematology"[All Fields]) AND ("flow cytometry"[All Fields] AND "digital imaging"[All Fields] OR "artificial intelligence"[All Fields] OR "machine learning"[All Fields] OR "point-of-care testing"[All Fields] OR "hematological disorders"[All Fields] AND ("precision diagnostics"[All Fields] OR "personalized medicine"[All Fields]) AND 2014/01/01:2024/12/31[Date - Publication] AND "English"[Language] |
| IEEE Xplore | ("hematology analyzer" OR "automated hematology") OR ("flow cytometry" AND "digital imaging" AND "artificial intelligence" OR "point-of-care testing" OR "hematological disorders") OR ("precision diagnostics" OR "personalized medicine") AND ("Publication Year":"2014-2024") AND ("Language":"English") |
| SCOPUS | (TITLE-ABS-KEY("hematology analyzer" OR "automated hematology") AND ("flow cytometry" OR "digital imaging" OR "artificial intelligence" OR "machine learning" OR "point-of-care testing" OR "hematological disorders" OR "precision diagnostics" OR "personalized medicine")) AND (PUBYEAR > 2013 AND PUBYEAR < 2025) AND (LIMIT-TO (LANGUAGE,"English")) |
| Google Scholar | ("hematology analyzer" OR "automated hematology") OR ("flow cytometry" AND "digital imaging" OR "artificial intelligence" OR "machine learning" OR "point-of-care testing" OR "hematological disorders" OR "precision diagnostics" OR "personalized medicine") AND ("since:2010") |

- Studies focusing on automated hematology analyzers used in clinical or research settings.
- Studies discussing advancements in technologies, such as AI/machine learning, flow cytometry, digital imaging, and point-of-care testing within hematology.
- Original research articles, review articles, and meta-analyses published in peer-reviewed journals.
- Studies published in English within the last decade (2014-2024).

### 2.3.2. Exclusion Criteria

The exclusion criteria were as follows:

- Studies focusing solely on manual hematological techniques or non-automated cell counting methods.
- Studies primarily focused on veterinary hematology or other non-human applications.
- Editorials, letters to the editor, conference abstracts (unless substantial data are provided), and non-peer-reviewed publications.
- Studies not available in English.

The screening process involved two stages. First, the titles and abstracts of the identified articles were independently screened to assess their relevance based on the inclusion and exclusion criteria, and any discrepancies were resolved. Subsequently, the full texts of the potentially eligible articles were retrieved and assessed using the same criteria. Figure 2 illustrates the study-selection process.

### 2.4. Content Analysis and Reporting

This review employed a narrative synthesis approach to analyze and synthesize data extracted from the included studies. Narrative synthesis was chosen because it is well suited for synthesizing diverse evidence, including qualitative and quantitative data, and for exploring complex relationships between different concepts. This method allows for a flexible and iterative approach to data analysis, enabling identification of key themes and patterns in the literature.

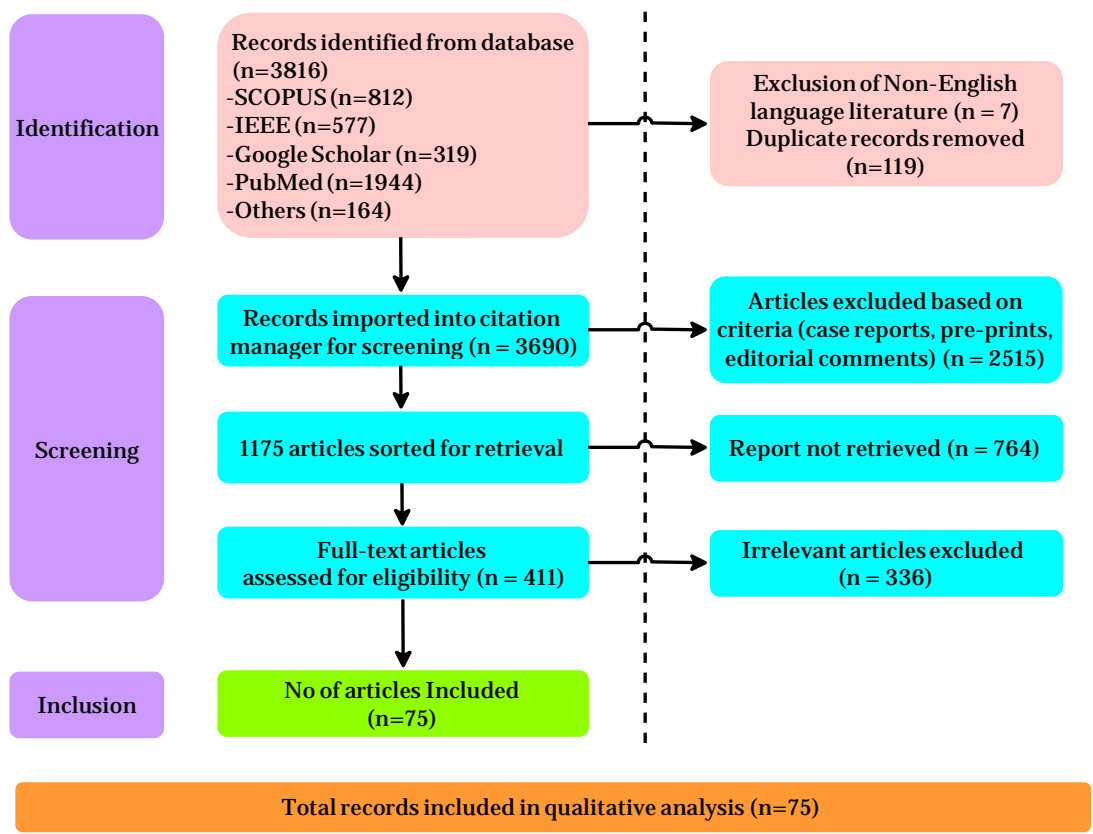

Figure 2.: Study selection process in PRISMA

No specific software was used for qualitative or quantitative data analysis, as the synthesis primarily focused on summarizing and interpreting the findings from the included studies. Data extraction involved summarizing key information from each study, including the study design, sample characteristics, technology used, key findings, and clinical implications. The results were classified and synthesized based on the key advancements in hematology analyzers identified in the research questions given in Table 3.

Table 3.: Research Questions

| No. | Research Question |
| --- | --- |
| 1 | How are artificial intelligence and machine learning applied to hematological analysis? |
| 2 | What are the recent advancements in flow cytometry techniques for hematological applications? |
| 3 | What improvements have been made in digital imaging technologies used in hematology? |
| 4 | What are the key developments in point-of-care testing of hematological parameters? |

Within each of these categories, findings were further synthesized to address specific aspects such as improvements in diagnostic accuracy, early disease detection, personalized treatment strategies, and impact on laboratory workflow. This structured approach facilitated a clear and organized presentation of the findings, allowing for a comprehensive overview of the current state of the field.

### 2.5. Bibliographic Analysis

This section analyzes the characteristics of the included studies to provide context and identify trends within the research on advancements in hematology analyzers. This analysis offers insights into the evolution of the field and highlights areas of focus in the literature.

#### 2.5.1. Publication Trends

The analysis of publication dates has revealed growing interest in this field over the past decade. The chart shows an upward trend, reflecting rapid technological advancements and research in hematology analysis. A bibliographic analysis of publications indexed in PubMed, Scopus, IEEE, Google Scholar, and other journals; conference proceedings; and even a few online available contents, using keywords related to hematology analyzers, revealed a significant increase in research output over the past decade (2014-2024) (Figure 3). This substantial growth in publications reflects the scientific community's increasing engagement in this field and underscores the growing importance of technological advancements in hematological diagnostics and personalized medicine, particularly in the post COVID-19 era.

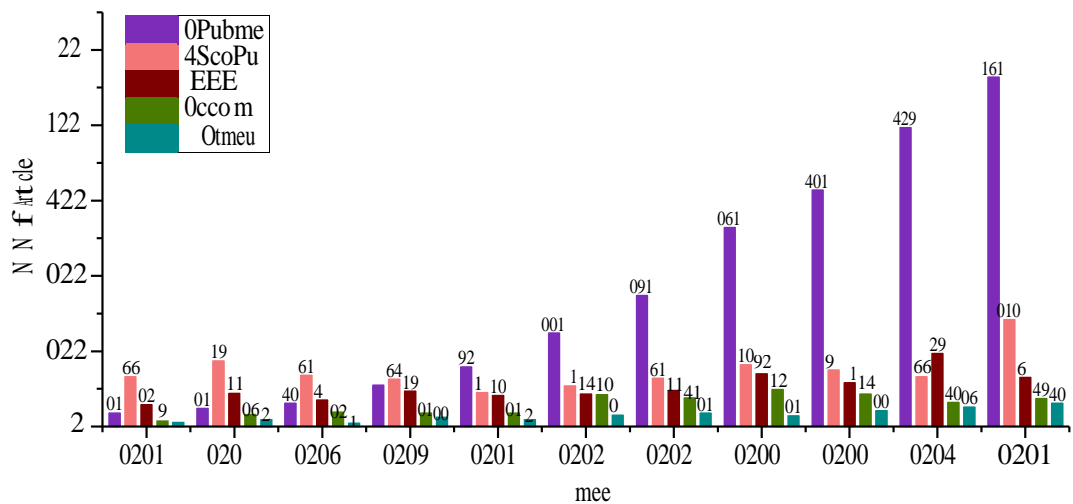

Figure 3.: Number of articles identified across databases through boolean search

### 2.5.2. Types of Studies

The included studies consisted of a mix of primary research articles (reporting original experimental data), review articles (synthesizing the existing literature), and potential meta-analyses (statistically combining the results from multiple studies). This mix provides a comprehensive view of the field, with primary research providing new data, and reviews offering broader perspectives.

### 2.5.3. Emerging Themes and Bibliometric Insights

Several key themes emerged from the bibliographic analysis.

- Increasing focus on AI and machine learning: A growing number of publications have explored the application of AI algorithms for automated cell identification, classification, and disease prediction in hematology.
- Advancements in flow cytometry for immunophenotyping and minimal residual disease detection: Research is increasingly focused on using flow cytometry for the more precise and sensitive detection of specific cell populations, particularly in the context of cancer diagnostics and monitoring.
- Integration of digital imaging with automated analysis: Studies have highlighted the benefits of combining high-resolution digital imaging with automated image analysis techniques for improved morphological assessment and reduced inter-observer variability.
- Growing interest in point-of-care testing: Recent publications emphasize the potential of POCT devices to improve access to hematological testing, particularly in resource-limited settings and for rapid diagnosis in emergency situations.

This bibliographic analysis provides a valuable overview of the research landscape in hematology analyzer technology, highlighting key trends and areas of active investigation. This sets the stage for a more detailed discussion of specific technological advancements and their clinical implications in the subsequent sections of the review.

Table 4.: Comparison of Traditional and Next-Generation Hematology Analyzers

| Feature | Traditional Analyzers | Next-Generation Analyzers |
|---|---|---|
| Applications | Routine hematology testing in smaller labs | Comprehensive hematology testing in medium to large labs, specialized hematology labs, research settings, and potentially point-of-care |
| Technology | Primarily impedance and optical light scattering | Advanced flow cytometry with digital imaging, AI analysis, and impedance |
| Analyzed Parameters | Limited set of cell counts (RBC, WBC, platelets) and differentials (neutrophils, lymphocytes, monocytes, eosinophils, basophils) | Expanded cell panel includes CBC, reticulocyte count, WBC differential with absolute counts, red and platelet indices, nucleated RBC count, immature granulocyte count, reticulocyte hemoglobin content, reticulocyte hemoglobin distribution width, along with functional assays and genetic analysis |
| Cell Analysis | Basic size and volume measurements | High-resolution morphology with flow cytometry, impedance, light scattering, and digital imaging |
| Data Analysis and Reporting | Manual or rule-based interpretation, basic reports, limited abnormal flagging | Advanced AI-powered automated analysis, machine learning algorithms, statistical tools, detailed graphical reports, and automated abnormal result flagging |
| Throughput | Moderate | High throughput with automation, parallel processing, and rapid turnaround time |
| Accuracy & Precision | Suitable for routine, limited in complex cases. | Enhanced accuracy and precision, especially for atypical cells and complex cases |
| Automation | Limited automation | High degree of automation, including sample preparation, analysis, and data management |
| Sample Volume | Relatively larger sample volume | Requires minimal sample (20–100 L) via microfluidics and sensitive detection |
| Cost | Lower initial cost, higher operational costs due to reagents and labor | Higher initial cost due to advanced technology, lower operational costs due to automation and reduced reagent consumption |
| Sensitivity and Specificity | Lower sensitivity and specificity for rare cells and subtle abnormalities | Higher sensitivity and specificity, enabling early detection of diseases and monitoring of treatment response |
| Portability | Limited portability, suited for central labs | Potential for POCT devices, enabling rapid diagnosis in remote or resource-limited settings |
| Complexity of Operation | Needs skilled technicians for operation and maintenance | Specialized training may be needed, but automation and user-friendly interfaces for ease of use |
| Integration with Other Systems | Limited integration with other laboratory systems | Potential for integration with LIS, EHRs, and other healthcare IT systems |
| Impact | Conventional analyzers offered valuable, basic blood counts | NGHAs are transforming hematology with enhanced precision, automation, and clinical insights, improving diagnosis and patient care |

## 3. Limitations of Conventional Hematology Analyzers: A Need for Deeper Insights

Although traditional hematology analyzers provide useful information concerning blood cell health, they cannot provide a complete picture of blood cell health. Some traditional basic analyzers provide a limited view of cell size, shape, and internal structure, all converging to reduce subtle changes that may be risk signals of early disease Kratz et al. (2019). In addition, determining cell function often requires a separate test, making the process much more time-consuming and complex Harrison (2005). Most importantly, traditional analyzers cannot explore the genetic background of blood cells, which is a determinant of many disorders related to blood Kratz and Brugnara (2015). These limitations restrict their application in precision medicine. In contrast, NGHAs are meant to fill this gap using modern techniques to improve blood cell health DeNicola (2011). Table 4 presents a general comparison between conventional hematology analyzers and NGHAs.

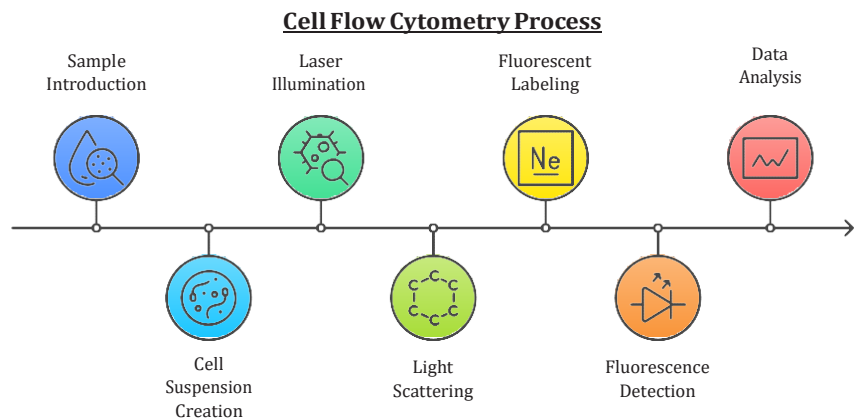

Figure 4.: Cell Flow Cytometry Process

## 4. Key Functionalities of Next-Generation Hematology Analyzers: Redefining Blood Analysis

NGHAs represent a major leap forward in blood analysis. Some significant features that set them apart from traditional analyzers are as follows.

### 4.1. Flow Cytometry

Flow cytometry is a new, rather revolutionary approach in hematology that has profoundly changed the way we think about examining cells in the blood. This advanced method is useful for observing each cell in a single file passing through a laser beam, thereby enabling the accurate identification and counting of diverse cell types. Figure 4 shows the process used for cell flow cytometry. Thus, it can be stated that flow cytometry is one of the key tools in the diagnosis of blood cancers, including leukemias and lymphomas Cools and Vandenberghe (2009) Engh Gerrit J. Van den (1989) Tomoyuki Kuroda (1991) Maekawa Yasunor Miki shi (1996). This helps doctors to accurately classify disease subtypes and track treatment progression. Additionally, flow cytometry plays a critical role in the stem cell transplantation process, enabling the successful engraftment and recovery of immune functions Vembadi et al. (2019) Robinson, Ostafe, Iyengar, Rajwa, and Fischer (2023) Orfao et al. (1995) Pang et al. (2023).

NGHAs help conduct blood examinations by offering a multiparameter count that provides a broad characterization of the cells in the blood. These include parameters, such as size, distribution, internal complexity, surface expression of antigens, granularity, and specific proteins. Figure 5 illustrates the basic working principle of the Cell Flow Cytometry system. Such a multidimensional cell analysis allows for a better understanding of cellular health. Such capabilities have far-reaching applications in hematology ranging from basic research to clinical diagnostics. For example, flow cytometry is widely used to distinguish and count different populations of blood cells such as lymphocytes, granulocytes, and monocytes. Such detailed cellular profiling can be highly useful in the diagnosis of many hematological disorders. Another ap-plication of flow cytometry is immunophenotyping Charles (1997), which is based on the concept of using antibodies specifically to differentiate cells based on the antigens expressed Heel et al. (2013) Marks et al. (2009). This has proven invaluable in characterizing cellular populations with high accuracy, and has been particularly useful in the diagnosis and classification of hematological malignancies. For example, in acute myeloid leukemia, flow cytometry can identify blast populations expressing CD34 and CD117 while also expressing CD13 markers for diagnostic and risk assessments Chisti (2006). In addition, flow cytometry helps evaluate residual disease and monitor minimal residual disease (MRD) Shumilov et al. (2018). This will be especially useful for diseases, such as leukemia, for which the presence of even the smallest number of surviving malignant cells could indicate the possibility of relapse. Advances in multi-color flow cytometry, which allows the simultaneous examination of many parameters along with advanced analysis software, have greatly enhanced the sensitivity of this technology. This makes complex cell populations more valuable for clinical studies and identification of rare cells better to identify. One of the most significant strengths of NGHAs is their ability to detect rare populations of cells, including circulating tumor cells and stem cells, which can have significant clinical implications. Moreover, flow cytometry has been used to identify and count many cell types in the immune system based on surface markers to diagnose immune disorders and monitor disease progression Robinson et al. (2023).

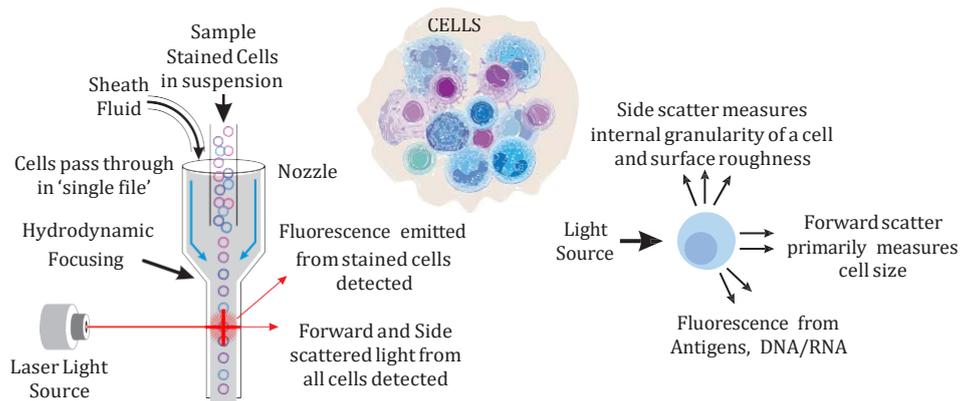

Figure 5.: Basic working principle of Cell Flow Cytometry

### *4.2. Advanced Imaging Techniques*

NGHAs have implemented more advanced imaging technologies, including flow cytometry, with new lasers and superior digital image resolution. Multicolor fluorescence identifies specific subpopulations with unique cell surface immune markers that can differentiate between complex leukemia and lymphoma diagnoses Cools and Vandenberghe (2009). These techniques make cell morphology more observable for details and thus allow for greater accuracy in revealing subtle abnormalities, such as when there is transformed cell morphology related to a given disease Tatsumi and Pierre (2002) Shean, Williams, and Rets (2024). Digital imaging techniques, such as fluorescence microscopy and confocal microscopy, enable the imaging of cellular morphology and interactions with unparallel details. Computer-assisted image analysis algorithms can be implemented to quantify various morphological parameters, identify abnormal cells, eliminate inter-observer variation, and enhance the quality of diagnosis. Digital images from blood films have become accessible and locatable for diagnosis from remote distances, thereby permitting telepathology and distance consultations.

### 4.3. Microfluidics Integration

Microfluidics is a revolutionary field in the field of blood analysis. By miniaturizing fluid handling systems, it allows for smaller, faster, and more efficient analyzers. Imagine tiny chips with intricate channels that can precisely manipulate minuscule blood samples. These microfluidic chips enable the integration of multiple tests, including functional analyses, on a single platform alongside traditional blood tests. This not only simplifies complex workflows, but also drastically reduces the amount of blood required for testing. By combining multiple tests on a single chip, microfluidics can provide a comprehensive blood analysis, offering valuable insights into patient health Kim, Zhbanov, and Yang (2022) Grigorev et al. (2023) Ding et al. (2023).

### 4.4. Artificial Intelligence Integration

Artificial intelligence (AI) has revolutionized the field of hematology. NGHAs generate massive amounts of data, and AI algorithms can analyze these data to identify subtle patterns and abnormalities that may be missed by human analysts. AI algorithms can classify abnormalities with remarkable accuracy by training on large datasets of blood cell images and associated pathologies, leading to earlier and more accurate diagnoses Fan et al. (2024) Walter et al. (2023). AI can also assist in automating data interpretation, eventually reducing human error, and improving diagnostic accuracy El Alaoui et al. (2022). AI-powered image analysis tools can analyze microscopic images of blood cells with greater precision, aiding in the detection of subtle morphological abnormalities. Additionally, AI can automate the classification of white blood cells into different subtypes, thereby reducing labor-intensive manual procedures and improving process efficiency. By analyzing large datasets, AI algorithms can identify patterns and trends, enabling early detection of diseases and personalized treatment strategies. This predictive capability of AI has immense potential to improve patient outcomes.

Figure 6 shows a block diagram of NGHA. The major components are as follows:

- Sample Inlet: This section represents the entry point for the blood sample.
- Sample Processing: The sample was processed (diluted/lysed etc.).
- Microfluidic Chip: This component illustrates the microfluidic channels responsible for the sample manipulation and analysis.
- Flow Cytometry Unit: This section houses the laser and detectors for analyzing the cell size, granularity, and fluorescence properties.
- Advanced Optics System: This block represents an imaging system for high-resolution cell morphology analysis.
- Integrated Biosensors: This section describes potential biosensors for the on-chip detection of specific biomarkers.
- AI Processing Unit: This block represents the electronics for capturing, analyzing, interpreting and processing data from various sensors along with AI integration with the system.

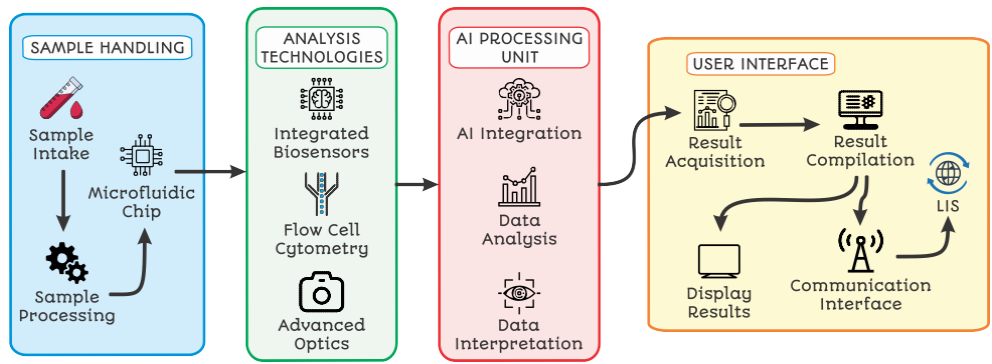

Figure 6.: Block Diagram of NGHAs Architecture

- User Interface: This section houses the computer responsible for data acquisition, and result generation and sends the result to the output media.
- Communication Interface: This block represents the connection for data transfer to laboratory information systems (LIS).

## 5. Integrating Cutting-Edge Technologies: A Synergistic Approach to Blood Analysis

NGHAs are testaments to the integration of cutting-edge technologies. Microfluidics revolutionizes blood analysis by miniaturizing fluid-handling systems Altendorf et al. (1998). Imagine tiny chips with intricate channels that can precisely manipulate minuscule blood samples. Heikali and Carlo (2010). This technology enables the integration of multiple tests such as functional analyses and traditional blood tests into a single chip. This not only streamlines complex workflows, but also significantly reduces the amount of blood required for testing, making it a more patient-friendly approach to screening. Flow cytometry is a powerful technique that uses lasers to analyze individual cells as they flow through a narrow stream Rosenbluth, Lam, and Fletcher (2008). Advanced flow cytometers in NGHAs employ sophisticated lasers and detectors to provide detailed information regarding the cell size, granularity, and surface markers. These data help differentiate between healthy and abnormal cells, even those with subtle differences, making them a valuable tool for diagnosing various blood disorders Drouet and Lees (1993) McCurley and Larson (1996). Microfluidics and flow cytometry, when combined, are potent analytical tools for in-depth blood cell examinations. Microfluidic technology drastically reduces the number of samples and related fluids required, making it possible to detect cellular irregularities efficiently and sensitively with minimal patient discomfort. This synergistic approach, paired with the high-throughput capabilities of flow cytometry, allows for swift multiparameter analysis of individual cells, providing invaluable insights into blood cell health and disease conditions Béné (2017). Nanoparticles engineered at the microscopic scale can be functionalized with specific molecules to target certain cellular features. In NGHAs, nanoparticles can be used as probes to bind to specific antigens on the surface of blood cells. These targeted probes enhance the sensitivity and specificity of NGHAs in detecting abnormalities, particularly in rare cell populations or subtle changes in antigen expression. This tech-nology holds great promise for early disease detection and personalized medicine Yao et al. (2020) Wu et al. (2022).

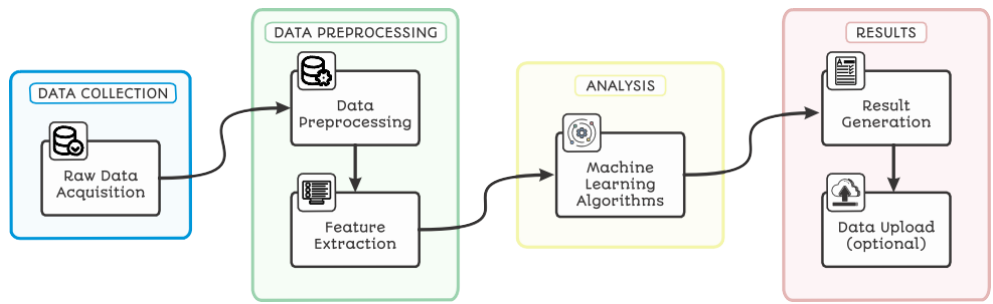

Figure 7.: Data analysis workflow in NGHAs

## 5.1. Data Analysis Workflow

The flowchart in Figure 7 outlines the steps involved in data analysis within a next-generation hematology analyzer.

(1) Raw Data Acquisition: In this process, the data was collected through different sensors; these include cell counts, size measurements, fluorescence intensity values, and possibly responses from biosensors.
(2) Data Preprocessing: This stage is the cleaning and normalization of raw data to assure accuracy and consistency.
(3) Feature Extraction: Major features derived from the preprocessed data such as cell size, granularity, fluorescence profiles, and possible biomarker signals were extracted.
(4) Machine Learning Algorithms: Artificial Intelligence algorithms are applied to analyze the extracted features, identify patterns, classify cell types, or detect anomalies.
(5) Result Generation: The system then generates a comprehensive report containing cell counts, differentials, morphology, and potential biomarker findings.
(6) Data Upload (Optional): The report and anonymized data can be uploaded to a Laboratory Information System for further analysis and integration into the patient record.

## 6. Considerations for Precision Diagnostics

To revolutionize hematology, next-generation analyzers must be designed using precision diagnostics. Conventional complete blood counts (CBCs) provide a basic snapshot of blood health. However, they often fail to detect subtle abnormalities or underlying diseases Chhabra (2018). Advanced analyzers should be capable of analyzing a wider range of cell types, including reticulocytes, immature granulocytes, platelets, and rare cell populations. This comprehensive approach provides a more detailed picture of blood cell health, allowing the identification of abnormal cell populations associated with hematological malignancies and other blood disorders Daves, Roccaforte, Lombardi, Panella, and Pastori (2023).

Table 5.: Examples of Potential Biomarkers Detectable by Next-Generation Hematology Analyzers

| Biomarker Category | Biomarker Examples | Disease Association |
|---|---|---|
| Genetic Mutations | FLT3-ITD, NPM1 mutations, CEBPA mutations, TP53 mutations, RAS, BCR-ABL1 fusion gene, PML-RAR fusion gene, JAK2 V617F mutation, CALR mutations, SF3B1 mutations, U2AF1 mutations | Acute myeloid leukemia (AML), chronic myeloid leukemia (CML), acute promyelocytic leukemia (APL), polycythemia vera, essential thrombocythemia, primary myelofibrosis, myelodysplastic syndromes (MDS) |
| Immune Activation | Elevated levels of inflammatory cytokines (IL-6, TNF-α, IFN-γ), increased expression of activation markers on immune cells (CD69, HLA-DR), alterations in T cell receptor repertoire | Autoimmune diseases, inflammatory bowel disease, rheumatoid arthritis, immune activation disorders, graft-versus-host disease |
| Cell Surface Markers | CD34+ hematopoietic stem cells, CD45+ leukocytes, CD14+ monocytes, CD15+ granulocytes, CD19+ B cells, CD3+ T cells, CD4+ T cells, CD8+ T cells | Leukemia, lymphoma, myelodysplastic syndromes, leukemias, lymphomas, infectious diseases, B-cell lymphomas, chronic lymphocytic leukemia, T-cell lymphomas, autoimmune diseases |
| MicroRNA Expression | miR-155, miR-125b, miR-21, miR-223, miR-146a | Hematological malignancies, autoimmune diseases, cardiovascular disease |
| Epigenetic Markers | DNA methylation, histone modifications (H3K27me3, H3K4me3, H3K9me3) | Cancer, aging, neurodegenerative diseases |
| Protein Expression | CD34, CD38, CD45RA, CD27, Ki-67 | Aging, immune senescence, cancer |
| Metabolic Markers | Lactate dehydrogenase (LDH), soluble CD25 (sCD25) | Cancer, inflammation |

Understanding these cell functions requires more than just cell counting. Advanced analytical tools are required to assess cellular functions, including oxidative stress, apoptosis, and phagocytosis. In addition, assessment of immune cells, such as T and B lymphocytes, can provide a substantial understanding of the body's immune response to diseases. The therapeutic response follow-up, particularly in patients with hematological malignancies, is another important aspect of functional assessment Rosales and Uribe-Querol (2017). Thus, the integration of on-chip DNA and RNA analysis technologies can be helpful in the discovery of genetic mutations and chromosomal abnormalities related to hematological diseases, such as leukemia, lymphoma, and hemophilia. Identification of genetic markers may predict susceptibility to diseases or response to treatment as a source of innovation in a discipline called personalized medicine. Table 5 presents a list of potential biomarkers that can be detected by NG-HAs. Tailoring therapeutic interventions according to an individual's genetic makeup can improve patient outcomes and reduce toxicity Z. Li, Xu, Wang, and Jiang (2023) Cho (2024). This integration of advanced functionalities into next-generation hematology analyzers significantly improves the accuracy and efficiency of blood tests, which helps in the earlier detection of conditions, better monitoring, and improved patient care.

## 7. Data Management and Standardization

Only an efficient next-generation hematology analyzer with effective management and standardization of data can bring out its full potential. The LIS interface is also important for ensuring streamlined workflow and data tracking, which are crucial for further streamlining data transfer, minimizing errors, and timely reporting Hulsen et al. (2019). Standardization is another important feature of data management. The use of standardized formats and reporting structures in data formats allows laboratories to interoperate between analyzers and different systems, making it easy to share, compare, and analyze data from other healthcare settings Panteghini and Forest (2005). Cloud-based data storage is a robust solution for the large amounts of data generated by NGHAs. Cloud platforms are secure, scalable, and accessible, allowing healthcare professionals remote access and collaboration. Furthermore, analytics tools can be applied to the cloud to extract valuable insights from data, which leads to improved decision making and patient care Banimfreg (2023) M. Li, Yu, Ren, and Lou (2010). To maintain data accuracy and reliability, robust quality control measures must be implemented to ensure data quality and safety. Patient information must be protected by encryption, access control, and periodic security audits M. Li et al. (2010) Benaloh, Chase, Horvitz, and Lauter (2009) Xhafa et al. (2015) Kumara, Nanumura, and Dissanayake (2023). Sophisticated analytical methodologies, including machine learning and artificial intelligence, can be used to derive significant insights from intricate data collection. Accessible visualization instruments can assist researchers and clinicians in efficiently comprehending and interpreting data Sharma, Lysenko, Jia, Boroevich, and Tsunoda (2024) Raparthi et al. (2021) Ahmed, Mohamed, Zeeshan, and Dong (2020). By addressing these essential components of data governance and standardization, it is possible to realize the complete capabilities of next-generation hematology analyzers and propel progress in personalized medicine.

## 8. Regulatory Considerations

The processes for regulatory clearance are vital for verifying the safety and efficiency of advanced hematology analyzers. Thorough validation studies are crucial to ensure the accuracy, precision, and reliability of new analytical methodologies. Such validation studies involved comparisons with established reference methods, reproducibility testing, and the detection of slight abnormalities. There is no method to determine the clinical feasibility of next-generation analyzers other than testing in a clinical setting. Such trials would include diversely recruited patients and a head-to-head comparison between the performance of new analyzers and standard techniques. Key outcomes include diagnostic accuracy, sensitivity, specificity, and clinical utility. These devices must comply with global regulatory frameworks such as ISO 13485 and the FDA to ensure their quality and safety Standardization (2016). Regulatory authorities thoroughly test the design, manufacturing processes, and performance metrics of next-generation hematology analyzers to ensure that they meet the stringent quality criteria.

## 9. Ethical Considerations

In addition to the substantial potential of NGHAs, it is necessary to consider significant ethical implications. Preservation of data privacy and security is of utmost importance, particularly in contexts involving sensitive patient information. Comprehensive data protection strategies must be established to protect personal data and prevent unauthorized access. The use of AI algorithms in NGHAs raises concerns regarding the potential for bias in data and algorithms. Therefore, it is crucial to develop and implement unbiased algorithms in order to prevent biased outcomes. Fair and equal access to these complex technologies represents a significant issue, because ultimately and only so will all individuals benefit from them, irrespective of their background and socio-economic status. The benefits of maximizing healthcare delivery through NGHAs with greater precision and productivity are still under debate and subject to varying interpretations of how these effects change the doctor-patient relationship. Human interaction is important for health care providers and patients; therefore, these measures must not be overlooked. We can examine some ethical considerations to ensure that the introduction of NGHAs is ethically aligned with societal benefits, as mentioned below.

### 9.1. Data Privacy And Security

Next-generation hematology analyzers generate a large amount of sensitive patient data, including genetic information, which require strict protection measures Ienca et al. (2018) Mittelstadt and Floridi (2016). Acquiring informed consent from patients is important. Patients must be fully informed about the collection, storage, and use of their data, and the right to withdraw participation in data sharing must not be violated Takei, Yokoyama, Yusa, Nakamura, and Ogawa (2018) Wang et al. (2024). Data must be anonymized, which involves removing all identifiers in the dataset such as names, addresses, and social security numbers. Data are also protected through various means, such as strong encryption methods, firewalls, and full-scale security audits, so that unauthorized access can be blocked along with cyber-attacks D'amico, Dall'Olio, Sala, Dall'Olio, and Sauta (2023) Cucoranu et al. (2013). Emphasizing data security, privacy, and ethical considerations would bring further benefits to employing the full potential of modern hematology analyzers in patient care, protecting specific rights, and preserving social value.

### 9.2. Bias in AI Algorithms

Moreover, AI-based algorithms are prone to inherent biases in the datasets used for training. Such algorithms may therefore provide incorrect diagnoses and treatment alternatives in some cases, and may not provide uniform treatment for all patients, especially minority populations Kocak et al. (2024) Mittermaier, Raza, and Kvedar (2023) Nazer, Zatarah, Waldrip, Ke, and Moukheiber (2023). This can be avoided by using mixed and representative datasets for training the AI algorithms, ensuring that the algorithms are sufficiently strong to correctly identify patterns over different patient groups. Transparency is also fundamental. Healthcare providers must understand how AI can reach such conclusions. Such transparency can help create trust between healthcare providers and their patients simultaneously, helping identify and correct biased errors. Although the hematology analyzers that use artificial intelligence possess great potential advantages, it should be remembered that they are not supposed to replace human knowledge and expertise. Healthcare providers must always participate in the interpretation of results and in decision-making processes. Artificial intelligence should be considered as a tool that can complement human capabilities and not replace them. Based on these findings, it can be assumed that AI-driven hematology analyzers are used in an ethical and responsible manner to improve the accuracy and equity of patient care.

### 9.3. Access To Healthcare

Next-generation hematology analyzers show great promise, but their advantages can only be realized when there is wider accessibility. Although these technologies have many advantages, access is still a sticking point Organization (2024). Cost considerations are a major barrier to scaling these technologies. Because developing and rolling out high-tech analyzers is expensive, their access is severely limited, particularly in resource-constrained settings. Efforts to reduce costs such as cheaper technology development and new financing models are important for ensuring equitable access. Global health initiatives and collaborations among international organizations, governments, and the private sector could bridge the gap in access to these technologies. Sharing knowledge, resources, and expertise may ensure the development and distribution of cost-effective next-generation hematology analyzers in resource-limited areas. Access to next-generation analyzers depends on the basic healthcare infrastructure, including trained personnel to operate the equipment and interpret the results, which requires a well-developed system. Prioritizing infrastructure development and capacity building in underresourced areas is essential for ensuring equitable access. Thus, better global health outcomes may be achievable if these barriers can be overcome and equitable access is assured to harness the full potential of next-generation hematology analyzers.

### 9.4. The Doctor-Patient Relationship

The introduction of next-generation hematology analyzers with AI-powered diagnostics has improved the relationship between doctors and patients to a better level. Although AI can enhance diagnostic accuracy and efficiency, the human connection between healthcare providers and patients must be sustained and well maintained. Building trust is paramount. Patients must trust the expertise of their respective health care providers. AI must be presented in such ways as conveying how it will facilitate doctors and not replacing them Schultze (2010)Asan, Bayrak, and Choudhury (2020). Open communication is essential in this regard. Doctors with clear explanations of the reason for the diagnosis will communicate with members of the patient's family about the outcomes of the AI-directed test. In this regard, patients build their understanding and clinical decision-making concerns are shared. Autonomy in healthcare decision-making should be awarded to patients. AI guidelines should evolve into recommendations rather than authority. Open communication is at the forefront of interest-based languages. Providers should converse with the patients regarding this process. Their character should help them to understand the add-ons and external limitations of such AI-oriented diagnostic decisions, which will lead to the resolution of diverse worries and inquiries. Moreover, it is equally important to recognize the psychological effects of AI on patients' mindset. AI may well supply anchoring facts and data-oriented suggestions, but will not touch upon the empathy and compassion that a physician has. Therefore, if a balance of AI alongside an infusion of humans is achieved, the discourse on positive patient experiences can find a receptive ear. Thoughtful consideration of these factors unlocks the potential for AI to improve patient care while simultaneously protecting the irreplaceable human element of the doctor-patient relationship.

### 9.5. Regulatory Frameworks

Regulatory frameworks are important drivers for the assurance of safety, efficiency, and ethics in the use of advanced hematology analyzers. Regulatory authorities include the United States Food and Drug Administration (FDA), Medical Device Regulation (MDR), and In-Vitro Diagnostic Regulation (IVDR) of the European Union, which have set strict criteria for the design development and clinical validation of devices Administration (2024) MDR (2024) IVDR (2024). To prove the accuracy, precision, and reliability of the new analytical technologies, extensive validation studies are required. These studies consisted of performance characterization by comparing the new analyzer with established reference techniques, examining its capability for reproducibility, and assessing its ability to identify subtle abnormalities. Clinical studies are essential for clinical validation of the utility of next-generation hematology analyzers in realistic applications. This study registered a large number of patients and evaluated the performance of the new analyzer compared to conventional methods. Data governance is an essential element of regulatory oversight mechanisms. There must be clear definitions and rule sets that define the ownership, storage, and utilization of data. This includes safeguarding patient anonymity and minimizing the risk of data leakage. For example, the EU's General Data Protection Regulation (GDPR) can serve as a model for implementing data governance, particularly in healthcare of the European Union (2024). Finally, rigorous validation processes are necessary for the application of AI algorithms in hematological diagnostics. Regulatory bodies can collaborate with research institutes and industry partners to formulate clear-cut rules and standards for the development, testing, and implementation of AI-driven hematology analyzers. Such steps are instrumental in ensuring the reliability, interpretability, and ethical use of these algorithms. Strong regulatory frameworks would bring about the responsible development and deployment of next-generation analyzers for hematology, a boon that benefits patients and their healthcare providers everywhere.

## 10. Impact On Personalized Medicine: Revolutionizing Blood-Based Diagnostics and Treatment

The next generation of hematology analyzers has revolutionized the domain of personalized medicine with more in-depth and detailed analyses of blood cells. This has been possible because of the inclusion of advanced features such as early diagnosis, AI-based risk stratification Elsabagh et al. (2023), targeted therapy, effective monitoring of treatment, liquid biopsy, rare disease diagnosis, and monitoring stem cell transplantation, which have transformed the NGHAs approach towards blood-related disorders. Figure 8 shows the general impact of the NGHAs on the personalized medicine paradigm.

### 10.1. Early Disease Detection

NGHAs can detect minute anomalies in blood cells that may show early signs of diseases such as leukemia, lymphoma, and other blood-related cancers. Early identification of these issues helps doctors to act quickly and provide better treatment, leading to better patient outcomes.

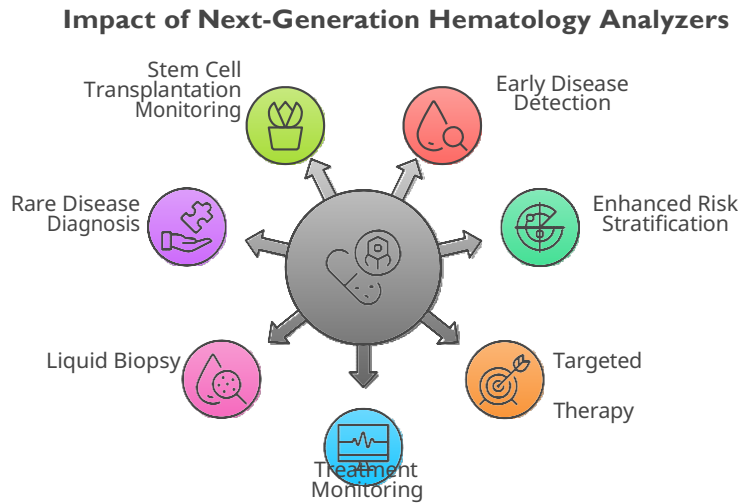

Figure 8.: NGHAs impact on personalized medicine

## 10.2. Better Risk Assessment

By examining many different biomarkers, NGHAs can be used to identify individuals who are more likely to develop certain diseases. This information can help put specific preventive steps in place, such as changes in lifestyle or early testing, to lower the chances of becoming sick.

## 10.3. Targeted Therapy

NGHAs can provide useful information regarding the genes and molecules involved in tumors. This information can guide the selection of targeted therapies based on the specific cause of the disease. Such targeted treatments are most likely to provide better results, with fewer harmful side effects.

## 10.4. Monitoring Treatment

NGHAs can help monitor the response of patients to therapy by tracking changes in blood cell measurements. This helps health care providers see how well the treatment is working, identify any possible side effects, and change their treatment plans if necessary.

## 10.5. Liquid biopsy

NGHAs allow the scrutiny of circulating tumor cells (CTCs) and circulating tumor DNA (ctDNA) in the whole blood. Therefore, this noninvasive technique will provide important information on tumor burden, rate of disease progression, and drug responsiveness.

### 10.6. Rare Disease Diagnosis

NGHAs may also be used to identify particular genetic mutations or protein abnormalities that contribute to the development of rare blood disorders, with earlier diagnosis followed by better management.

### 10.7. Stem Cell Transplantation Monitoring

The use of NGHAs may be beneficial for monitoring engraftment and recovery of transplanted stem cells. Such optimization may avoid treatment-related complications. By integrating these advanced functionalities, next-generation hematology analyzers will power healthcare providers to treat patients with greater precision, personalization, and, ultimately, even better patient care.

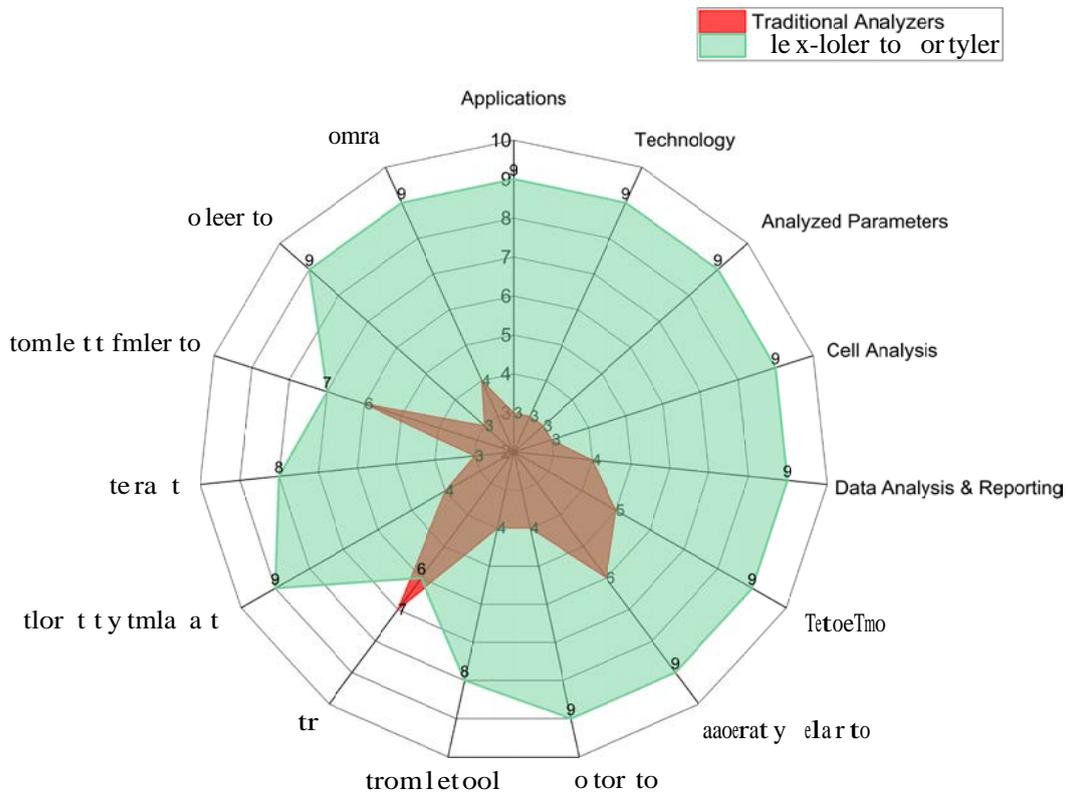

Figure 9.: Radar chart highlighting the transformative impact of NGHAs compared to the conventional hematology analyzer

Figure 9 illustrates a radar chart comparing the capabilities of traditional hematology analyzers (TAs) and next-generation hematology analyzers (NGHAs) across 15 key performance indicators on a scale of 1 to 10. This chart clearly demonstrates the transformative potential of NGHAs for advancing personalized medicine. The significantly larger area covered by the NGHA data series highlights their superiority in critical aspects such as comprehensive cell analysis, advanced data reporting, and high-throughput capabilities with enhanced accuracy and precision. It is of particular importance that the chart demonstrates the superior ability of NGHAs in measuring a wider range of parameters with enhanced sensitivity and specificity, enabling a deeper understanding of individual patient profiles at the cellular and molecular levels. This detailed and individualized analysis, facilitated by NGHA technology, is fundamental for tailoring diagnostics and treatment strategies, marking a significant leap toward truly personalized medicine in hematology.

## 11. Challenges and Future Directions: Pushing The Boundaries of Hematology Analysis

Although Next-Generation Hematology Analyzers (NGHAs) hold immense promise, hurdles still need to be overcome. Table 6 highlights the key strategies that NGHAs should implement to enhance their capabilities and performance further. A significant challenge is the high cost and accessibility of advanced devices. It is crucial to make them affordable and accessible to health care facilities worldwide Gurkan et al. (2024). Another challenge lies in managing and interpreting the vast amounts of complex data generated by the NGHAs. Developing user-friendly interfaces and robust data management systems are essential for their widespread adoption Tolan et al. (2015). Standardized protocols and rigorous validation processes are vital for maintaining the accuracy and reliability of results across different laboratories Verbrugge and Huisman (2015). Therefore, the future of NGHAs is exciting. The integration of microfluidics and point-of-care testing technologies is poised to revolutionize healthcare Ding et al. (2023) Choi (2016) Baratchi et al. (2014) F. Li et al. (2020). These advances can lead to rapid on-site diagnostics, particularly in resource-limited settings Yager et al. (2006). The development of portable and user-friendly NGHAs suitable for point-of-care testing in clinical settings or home environments holds immense promise for rapid diagnostics and personalized healthcare management Bagnall (2024) Bransky, Larsson, Aardal, Ben-Yosef, and Christenson (2021) Rao et al. (2008). Furthermore, combining hematological analysis with other diagnostic assays on a single chip, known as lab-on-a-chip technology, can provide a comprehensive health profile from a minimal blood sample Gurkan et al. (2024). Single-cell analysis, enabled by advancements in microfluidics, offers the potential to delve deeper into the intricacies of individual blood cells, thereby providing valuable insights into the cellular health and heterogeneity within cell populations Brierley, Charlotte, and Mead; Adam (2020). Finally, the integration of machine learning and AI algorithms may help improve diagnosis using NGHAs. Self-learning systems can evaluate blood cell patterns, and more precise and accurate diagnoses can be achieved Muhsen, Shyr, Sung, and Hashmi (2020) El Alaoui et al. (2022). Advances in technology will thus enable NGHAs to take center stage and herald a revolution in hematology toward earlier detection, more precise diagnosis, and tailor-made strategies for treatment.

## 12. Conclusion

Next-generation hematology analyzers are the most significant breakthroughs in the history of blood analysis. The inclusion of sophisticated technologies, such as microfluidics, flow cytometry, and artificial intelligence, has provided NGHAs with an expanded array of cellular parameters, improved sensitivity, and the possibility of on-chip amalgamation of several assays. These instruments will revolutionize personalized medical strategies in the biomedical engineering domain. Nonetheless, tackling the is-sues associated with data management, standardization, and ethical considerations is essential for effective implementation of these analyzers. As this domain progresses, next-generation hematology analyzers can revolutionize healthcare delivery, resulting in earlier diagnosis, targeted therapies, and enhanced patient outcomes. Although the challenges of cost, data management, and standardization remain, NGHAs promise to revolutionize blood-based diagnostics and personalized treatment strategies in the near future. By addressing these challenges and continuing innovation, we will unlock the full potential of NGHAs in improving patient care and saving lives.

Table 6.: Future Research Directions for Hematology Analyzers

| Research Direction | Description | Potential Impact |
|---|---|---|
| Integration of Multi-omics Data with Hematology Analyzers | Developing new computer programs combining information from genomics, proteomics, metabolomics, and all other "omics" technologies with information drawn from blood tests to provide an improved view of patient health by identifying novel biomarkers or estimating disease risk Reel, Reel, Pearson, Trucco, and Jefferson (2021). | Improved understanding of disease mechanisms, identification of new diagnostic and prognostic markers, development of more targeted therapies, and more accurate risk stratification. |
| Development of Advanced AI Algorithms for Complex Morphology Analysis | Creating better AI and deep learning programs that can study complex cell shapes more accurately. This might include training these programs using large collections of clear images to identify small morphological anomalies that humans may overlook. | Earlier and more accurate diagnosis of hematological malignancies and other disorders reduces interobserver variability in morphological assessments and improves the efficiency of laboratory workflows. |
| Miniaturization and Integration of Microfluidics for Point-of-Care Testing | Small, integrated microfluidic devices that can be developed for "lab-on-a-chip" will allow blood tests to be carried out fast and cheaply at point-of-care sites Baratchi et al. (2014) Luppa, Bietenbeck, Beaudoin, and Giannetti (2016) Cummins, Ligler, and Walker (2016). Such devices perform a series of tests on an extremely small amount of blood and enable rapid diagnosis, even in resource-poor settings or at the patient's bedside. | Improved access to hematological testing, especially in underdeveloped regions and developing countries; accelerated processing times for critical tests; maximized patient care in emergency situations; and reduced healthcare costs. |
| Development of Novel Biomarkers for Early Disease Detection | The focus shall be on the creation of new markers identifiable by hematology analyzers, thus aiding in the early diagnosis of hematological disorders, particularly malignancies. This encompasses testing for circulating tumor cells, cell-free DNA, and other molecular markers. | Early disease detection results in enhanced treatment outcomes and increased survival probability. Moreover, it facilitates prompt execution of preventative strategies. |
| Standardizing Data and Interoperability of Hematology Analyzers | Setting up common data formats and communication methods for blood testing analyzers to help share and combine data between different labs and healthcare systems. This could allow for the development of large, multi-center datasets for research and the creation of new diagnostic tools. | This will enable stronger algorithms in diagnosis and tailored treatment plans through collaboration and information sharing between researchers and clinicians. This new development further supports quality control and reproducibility of results in laboratories worldwide. |

# References

Administration, U. F. . D. (2024). *Overview of device regulation.* Retrieved 2024-11-15, from `https://www.fda.gov/medical-devices/`

Ahmed, Z., Mohamed, K., Zeeshan, S., & Dong, X. (2020). Artificial intelligence with multi-functional machine learning platform development for better healthcare and precision medicine. *Database*, *2020*, baaa010.

Altendorf, E., Zebert, D., Holl, M., Vannelli, A., Wu, C., & Schulte, T. (1998). Results obtained using a prototype microfluidics-based hematology analyzer. In *Micro total analysis systems '98* (p. 73-76). Springer Netherlands.

Asan, O., Bayrak, A. E., & Choudhury, A. (2020). Artificial intelligence and human trust in healthcare: Focus on clinicians. *Journal of Medical Internet Research*, *22*(6).

Bagnall, D. M. R. L. B. I. N. A. F. G. B. L. I. O. O. I., Robert; Guy. (2024). Point-of-care diagnostic test accuracy in children and adolescents with sickle cell disease: A systematic review and meta-analysis. *Blood Reviews*, 101243.

Banimfreg, B. H. (2023). A comprehensive review and conceptual framework for cloud computing adoption in bioinformatics. *Healthcare Analytics*, *3*, 100190.

Baratchi, S., Khoshmanesh, K., Sacristán, C., Depoil, D., Wlodkowic, D., McIntyre, P., & Mitchell, A. (2014). Immunology on chip: Promises and opportunities. *Biotechnology Advances*, *32*(2), 333-346.

Benaloh, J., Chase, M., Horvitz, E., & Lauter, K. (2009). Patient controlled encryption: ensuring privacy of electronic medical records. In *Proceedings of the 2009 acm workshop on cloud computing security* (p. 103–114). New York, NY, USA: Association for Computing Machinery.

Bransky, A., Larsson, A., Aardal, E., Ben-Yosef, Y., & Christenson, R. H. (2021). A novel approach to hematology testing at the point of care. *Journal of Applied Laboratory Medicine*, *6*(2), 532-542.

Brierley, Charlotte, K., & Mead; Adam, J. (2020). Single-cell sequencing in hematology. *Current Opinion in Oncology*, *32*(2), 139-145.

Béné, M. C. (2017). Microfluidics in flow cytometry and related techniques. *International Journal of Laboratory Hematology*, *39*(S1), 93-97.

Charles, C. (1997). Preserved cell preparations for flow cytometry and immunology. *Biotechnology Advances*, *15*(1), 100.

Chhabra, G. (2018, 01). Automated hematology analyzers: Recent trends and applications. *Journal of Laboratory Physicians*, *10*, 15.

Chisti, Y. (2006). Flow cytometry for speedy multifactor analyses in biotechnology. *Biotechnology Advances*, *24*(3), 352.

Cho, Y.-U. (2024). The role of next-generation sequencing in hematologic malignancies. *Blood Research*, *59*(11).

Choi, S. (2016). Powering point-of-care diagnostic devices. *Biotechnology Advances*, *34*(3), 321-330. (Trends in In Vitro Diagnostics and Mobile Healthcare)

Cools, J., & Vandenberghe, P. (2009). New flow cytometry in hematologic malignancies. *Haematologica*, *94*(12), 1639-1641.

Cucoranu, I. C., Parwani, A. V., West, A. J., Romero-Lauro, G., Nauman, K., Carter, A. B., ... Pantanowitz, L. X. (2013). Privacy and security of patient data in the pathology laboratory. *Journal of pathology informatics*, *4*(1), 4.

Cummins, B. M., Ligler, F. S., & Walker, G. M. (2016). Point-of-care diagnostics for niche applications. *Biotechnology Advances*, *34*(3), 161-176.

D'amico, S., Dall'Olio, D., Sala, C., Dall'Olio, L., & Sauta, E. (2023). Synthetic data generation by artificial intelligence to accelerate research and precision medicine in hematology. *JCO Clinical Cancer Informatics*, *7*, e2300021.

Daves, M., Roccaforte, V., Lombardi, F., Panella, R., & Pastori, S. (2023). Modern hematology analyzers: beyond the simple blood cells count (with focus on the red blood cells). *Journal of Laboratory and Precision Medicine*, *9*.

DeNicola, D. B. (2011). Advances in hematology analyzers. *Topics in Companion Animal Medicine*, *26*(2), 52-61.

Ding, L., Oh, S., Shrestha, J., Lam, A., Wang, Y., Radfar, P., & Warkiani, M. E. (2023). Scaling up stem cell production: harnessing the potential of microfluidic devices. *Biotechnology Advances*, *69*, 108271.

Drouet, M., & Lees, O. (1993). Clinical applications of flow cytometry in hematology and immunology. *Biology of the Cell*, *78*(1), 73-78.

El Alaoui, Y., Elomri, A., Qaraqe, M., Padmanabhan, R., Taha, R., Omri, H., ... Aboumarzouk, O. (2022). A review of artificial intelligence applications in hematology management: Current practices and future prospects. *Journal of Medical Internet Research*, *24*(7), e36490.

Elsabagh, A. A., Elhadary, M., Elsayed, B., Elshoeibi, A. M., Ferih, K., Kaddoura, R., ... Yassin, M. (2023). Artificial intelligence in sickle disease. *Blood Reviews*, *61*, 101102.

Engh Gerrit J. Van den, B. J. T. (1989). Detection of specific dna sequences by flow cytometry. *Biotechnology Advances*, *7*(2), 261.

Fan, B. E., Yong, B. S. J., Li, R., Wang, S. S. Y., Aw, M. Y. N., Chia, M. F., ... Winkler, S. (2024). From microscope to micropixels: A rapid review of artificial intelligence for the peripheral blood film. *Blood Reviews*, *64*, 101144.

Grigorev, G. V., Lebedev, A. V., Wang, X., Qian, X., Maksimov, G. V., & Lin, L. (2023). Advances in microfluidics for single red blood cell analysis. *Biosensors*, *13*(1), 117.

Gurkan, U. A., Wood, D. K., Carranza, D., Herbertson, L. H., Diamond, S. L., Du, E., ... Lam, W. A. (2024). Next generation microfluidics: fulfilling the promise of lab-on-a-chip technologies. *Lab Chip*, *24*(7), 1867-1874.

Harrison, P. (2005). Platelet function analysis. *Blood Reviews*, *19*(2), 111-123.

Heel, K., Tabone, T., Rohrig, K. J., Maslen, P. G., Meehan, K., Grimwade, L. F., & Erber, W. N. (2013). Developments in the immunophenotypic analysis of haematological malignancies. *Blood Reviews*, *27*(4), 193-207.

Heikali, D., & Carlo, D. D. (2010). A niche for microfluidics in portable hematology analyzers. *JALA: Journal of the Association for Laboratory Automation*, *15*(4), 319-328.

Hulsen, T., Jamuar, S. S., Moody, A. R., Karnes, J. H., Varga, O., Hedensted, S., ... McKinney, E. F. (2019). From big data to precision medicine. *Frontiers in Medicine*, *6*, 34.

Ienca, M., Ferretti, A., Hurst, S., Puhan, M., Lovis, C., & Vayena, E. (2018). Considerations for ethics review of big data health research: A scoping review. *PLoS One*, *13*(10).

IVDR, T. E. (2024). *The european union in vitro diagnostic regulation.* Retrieved 2024-10-20, from https://euivdr.com/

Kim, H., Zhbanov, A., & Yang, S. (2022). Microfluidic systems for blood and blood cell characterization. *Biosensors (Basel)*, *13*(1), 13.

Kocak, B., Ponsiglione, A., Stanzione, A., Blüthgen, C., Santinha, J., Ugga, L., ... Cuocolo, R. (2024). Bias in artificial intelligence for medical imaging: fundamentals, detection, avoidance, mitigation, challenges, ethics, and prospects. *Diagnostic and interventional radiology (Ankara, Turkey)*.

Kratz, A., & Brugnara, C. (2015). Automated hematology analyzers: State of the art. *Clin Lab Med*, *35*(1), 13.

Kratz, A., hee Lee, S., Zini, G., Riedl, J. A., Hur, M., & Machin, S. (2019). Digital morphology analyzers in hematology: Icsh review and recommendations. *International Journal of Laboratory Hematology*, *41*(4), 437-447.

Kumara, I. N., Nanumura, U. A., & Dissanayake, H. (2023). Enhancing data privacy of medical data through encryption and access control. *International Journal of Research in Engineering, Science and Management*, *6*(11), 38-43.

Li, F., You, M., Li, S., Hu, J., Liu, C., Gong, Y., ... Xu, F. (2020). Paper-based point-of-care immunoassays: Recent advances and emerging trends. *Biotechnology Advances*, *39*, 107442.

Li, M., Yu, S., Ren, K., & Lou, W. (2010). Securing personal health records in cloud computing: Patient-centric and fine-grained data access control in multi-owner settings. In *Security and privacy in communication networks* (Vol. 50, p. 89-106). Berlin, Heidelberg: Springer Berlin Heidelberg.

Li, Z., Xu, X., Wang, D., & Jiang, X. (2023). Recent advancements in nucleic acid detection with microfluidic chip for molecular diagnostics. *TrAC Trends in Analytical Chemistry*, *158*.

Luppa, P. B., Bietenbeck, A., Beaudoin, C., & Giannetti, A. (2016). Clinically relevant analytical techniques, organizational concepts for application and future perspectives of point-of-care testing. *Biotechnology Advances*, *34*(3), 139-160.

Maekawa Yasunor Miki shi, H. k. (1996). Method of classifying leukocytes by flow cytometry. *Biotechnology Advances*, *14*(4), 498-499.

Marks, D. I., Paietta, E. M., Moorman, A. V., Richards, S. M., Buck, G., DeWald, G., ... Lazarus, H. M. (2009). T-cell acute lymphoblastic leukemia in adults: clinical features, immunophenotype, cytogenetics, and outcome from the large randomized prospective trial (ukall xii/ecog 2993). *Blood*, *114*(25), 5136-5145.

McCurley, T. L., & Larson, R. (1996). Clinical applications of flow cytometry in hematology and immunology. *Clinical laboratory science: Journal of the American Society for Medical*

*Technology*, *9*(6), 358-362.

MDR, T. E. (2024). *The european union medical device regulation.* Retrieved 2024-12-1, from `https://eumdr.com/`

Mittelstadt, B., & Floridi, L. (2016). *The ethics of biomedical big data.* Imprint: Springer.

Mittermaier, M., Raza, M. M., & Kvedar, J. C. (2023). Bias in ai-based models for medical applications: challenges and mitigation strategies. *NPJ Digital Medicine*, *6*(1), 113.

Muhsen, I. N., Shyr, D., Sung, A. D., & Hashmi, S. K. (2020). Machine learning applications in the diagnosis of benign and malignant hematological diseases. *Clinical Hematology International*, *3*(1), 13-20.

Nazer, L. H., Zatarah, R., Waldrip, S., Ke, J. X. C., & Moukheiber, M. (2023). Bias in artificial intelligence algorithms and recommendations for mitigation. *PLOS Digital Health*, *2*(6).

of the European Union, H. . F. P. (2024). *Complete guide to gdpr compliance.* Retrieved 2024-1-7, from `https://gdpr.eu/`

Orfao, A., Ruiz-Arguelles, A., Lacombe, F., Ault, K., Basso, G., & Danova, M. (1995). Flow cytometry: its applications in hematology. *Haematologica*, *80*(1), 69-81.

Organization, W. H. (2024). *Access to medicines and health products.* Retrieved 2024-11-15, from `https://www.who.int/our-work/`

Pang, K., Dong, S., Zhu, Y., Zhu, X., Zhou, Q., Gu, W., Bobo; Jin, ... Wei, X. (2023). Advanced flow cytometry for biomedical applications. *Journal of Biophotonics*, *16*(9).

Panteghini, M., & Forest, J. C. (2005). Standardization in laboratory medicine: New challenges. *Clinica chimica acta; international journal of clinical chemistry*, *355*(1-2), 1-12.

Rao, L. V., Ekberg, B., Connor, D., Jakubiak, F., Vallaro, G., & Snyder, M. (2008). Evaluation of a new point of care automated complete blood count (cbc) analyzer in various clinical settings. *Clinica Chimica Acta; International Journal of Clinical Chemistry*, *389*(1-2), 120-125.

Raparthi, M., Kasaraneni, B. P., Kondapaka, K. K., Gayam, S. R., Thuniki, P., Kuna, S. S., ... Sahu, S. P. P. M. K. (2021). Precision health informatics-big data and ai for personalized healthcare solutions: Analyzing their roles in generating insights and facilitating personalized healthcare solutions. *Human-Computer Interaction Perspectives*, *1*(2), 1-8.

Reel, P. S., Reel, S., Pearson, E., Trucco, E., & Jefferson, E. (2021). Using machine learning approaches for multi-omics data analysis: A review. *Biotechnology Advances*, *49*, 107739.

Robinson, J. P., Ostafe, R., Iyengar, S. N., Rajwa, B., & Fischer, R. (2023). Flow cytometry: The next revolution. *Cells*, *12*(14), 1875.

Rosales, C., & Uribe-Querol, E. (2017). Phagocytosis: A fundamental process in immunity. *BioMed Research International*, 18.

Rosenbluth, M. J., Lam, W. A., & Fletcher, D. A. (2008). Analyzing cell mechanics in hematologic diseases with microfluidic biophysical flow cytometry. *Lab Chip*, *8*(7), 1062-1070.

Schultze, J. L. (2010). Building trust in medical use of artificial intelligence - the swarm learning principle. *Journal of CME*, *12*(1).

Sharma, A., Lysenko, A., Jia, S., Boroevich, K. A., & Tsunoda, T. (2024). Advances in ai and machine learning for predictive medicine. *Journal of Human Genetics*, *69*(10), 487-497.

Shean, R. C., Williams, M. C., & Rets, A. V. (2024). Advances in hematology analyzers technology. *Clinics in Laboratory Medicine*, *44*(3), 377-386.

Shumilov, E., Flach, J., Kohlmann, A., Banz, Y., Bonadies, N., Fiedler, M., ... Bacher, U. (2018). Current status and trends in the diagnostics of aml and mds. *Blood Reviews*, *32*(6), 508-519.

Standardization, I. O. (2016). *Iso 13485:2016 - medical devices - quality management systems - requirements for regulatory purposes.* Retrieved 2024-11-15, from `https://www.iso.org/standard/59752.html`

Takei, T., Yokoyama, K., Yusa, N., Nakamura, S., & Ogawa, M. (2018). Artificial intelligence guided precision medicine approach to hematological disease. *Blood*, *132*, 2254.

Tatsumi, N., & Pierre, R. V. (2002). Automated image processing: past, present, and future of blood cell morphology identification. *Clinics in laboratory medicine*, *22*(1), 299–315.

Tolan, N. V., Parnas, M. L., Baudhuin, L. M., Cervinski, M. A., Chan, A. S., Holmes, D. T., . . . Master, S. R. (2015). Big data" in laboratory medicine. *Clinical Chemistry*, *61*(12), 1433-1440.

Tomoyuki Kuroda, K. (1991). Reagent for reticulocyte counting by flow cytometry. *Biotechnology Advances*, *9*(2), 294.

Vembadi, A., Menachery, A., & Qasaimeh, M. A. (2019). Cell cytometry: Review and perspective on biotechnological advances. *Frontiers in bioengineering and biotechnology*, *7*, 147.

Verbrugge, S., & Huisman, A. (2015). Verification and standardization of blood cell counters for routine clinical laboratory tests. *Clinics in Laboratory Medicine*, *35*(1), 183-196.

Walter, W., Pohlkamp, C., Meggendorfer, M., Nadarajah, N., Kern, W., Haferlach, C., & Haferlach, T. (2023). Artificial intelligence in hematological diagnostics: Game changer or gadget? *Blood reviews*, *58*, 101019.

Wang, S.-X., Huang, Z.-F., Li, J., Wu, Y., Du, J., & Li, T. (2024). Optimization of diagnosis and treatment of hematological diseases via artificial intelligence. *Frontiers in Medicine*, *11*, 1487234.

Wu, P., Han, J., Gong, Y., Liu, C., Yu, H., & Xie, N. (2022). Nanoparticle-based drug delivery systems targeting tumor microenvironment for cancer immunotherapy resistance: Current advances and applications. *Pharmaceutics*, *14*(10), 1990.

Xhafa, F., Li, J., Zhao, G., Li, J., Chen, X., & Wong, D. S. (2015). Designing cloud-based electronic health record system with attribute-based encryption. *Multimedia Tools Appl.*, *74*(10), 3441–3458.

Yager, P., Edwards, T., Fu, E., Helton, K., Nelson, K., Tam, M. R., & Weigl, B. H. (2006). Microfluidic diagnostic technologies for global public health. *Nature*, *442*(7101), 412-418.

Yao, Y., Zhou, Y., Liu, L., Xu, Y., Chen, Q., & al., e. (2020). Nanoparticle-based drug delivery in cancer therapy and its role in overcoming drug resistance. *Frontiers in molecular biosciences*, *7*, 193.